\documentclass[12pt]{article}

\usepackage{scicite}

\usepackage{times}

\usepackage{graphicx}

\usepackage{CJKutf8} 

\newenvironment{sciabstract}{%
\begin{quote} \bf}
{\end{quote}}

\title{Preservation of the harmonic overtones in the violin family beyond Stradivari}

\author
{Li Su,$^{1}$ Chen-Chieh Chang,$^{2}$ Yu-Ming Lin,$^{3}$ Li-Chun Lu,$^{4}$ Yi-Wen Liu,$^{4}$\\ Ming-Sian Bai,$^{5}$ Chyh-Hong Chern$^{3,6,7\ast}$\\
\\
\normalsize{$^{1}$Institute of Information Science, Academia Sinica, Taipei 11529, Taiwan}\\
\normalsize{$^{2}$Institute of Education, National Taiwan Ocean University, Keelung 202301, Taiwan}\\
\normalsize{$^{3}$String Bean Ltd.,12F, No. 101, Section 2nd, Nanjing E. Road, Taipei 10457, Taiwan}\\
\normalsize{$^{4}$Department of Electrical Engineering, National Tsing-Hua University, Hsinchu 30013, Taiwan}\\
\normalsize{$^{5}$Department of Power Mechanical Engineering, National Tsing-Hua University, Hsinchu 30013, Taiwan}\\
\normalsize{$^{6}$Physics Division, National Center for Theoretical Sciences, Taipei 10617, Taiwan}\\
\normalsize{$^{7}$Department of Physics, National Taiwan University, Taipei 10617, Taiwan}\\
\\
\normalsize{$^\ast$To whom correspondence should be addressed; E-mail:  chern@alumni.stanford.edu.}
}

\date{}

\begin{document} 


\baselineskip24pt


\maketitle


\begin{sciabstract}
Three hundred years ago, Antonio Stradivari enjoyed the golden period of the violin making in the human history.  Luthiers and researchers endeavor to study his legendary legacy.  Unfortunately, a consensus that the present progress has reached his level remains lacking.  Most of the instruments suffer from low sounding power and enigmatic deficit of harmonic overtones.  In fact, there is significant energy dissipation in the acoustic couplings in the violin family, completely unaware to luthiers and researchers.  The current acoustic mechanism is inefficient in resonance of the wooden plates and hence is the low rate of mechanical energy conversion to the desired sound.  We propose a new method for the amelioration of the acoustic couplings to enhance the preservation of the overtone components up to 161\% of the pristine and to reduce the instrumental dissipation intensity up to 3.8 dB off (73\% off). It results in the tremendously improved sound projection, the shorter reaction time, the longer reverberation time, and consequently the enrichment of the timbre complexity.  Our physical method is applicable to the violin family of all sizes and ages, including the Stradivari, and extendable to all string instruments.
\end{sciabstract}


Music and the technology of musical instruments are one of the most important signatures in human civilization.  In particular, the making of violin family attracts tremendous interests in music, science, engineering, and education \cite{cremer,woodhouse2014}.  In brief summary of recent studies, the free plate tuning measures the resonance modes of the body plates before they are assembled \cite{hutchins1983}.  The shape and position of the renowned {\it f}-holes make violin family unique among all string instruments \cite{bissinger1992, nia2015}.  The admittance of the artistic bridge and the acoustic characterization of the resonance cavity have been actively studied \cite{jansson1997, bissinger1992, curtin2009, gough2005}.  Very different from all other scientific and engineering advances, the clock of progress in violin making seemingly forget to tick \cite{amadeus1998}.  Two ancient Italian luthiers, Antonio Stradivari and Guarneri del Ges\'u, remain the top benchmarks in the sound quality of all time.  In particular, their triumph is attributed to perceivable richness of the harmonic overtones, namely timbre complexity \cite{helmholtz1954}, dominant sound projection, and quality playability \cite{galluzzo2003}.  Their instruments are the most favorable to musical soloists, and their copies prevail the most popular models in the last three centuries.  The research to unveil their secrets spans an active field in music \cite{sacconi1979}, physics \cite{gough2000}, chemistry \cite{brandmair2010}, material science \cite{tai2017}, and acoustic engineering.

Musical instruments can be viewed as the physical apparatus that transform the biological energy from the players to the sound energy.  In the violin family, the strings and the resonance cavity are acoustically coupled by the accessories: pegs, a bridge, fine tuners, a tailpiece \cite{pigneret2018, stoppani2011}, a tailgut, and an endpin \cite{fleming2011}.   However, those complex acoustic couplings make the violin family highly dissipative.  A common and notorious problem among most of the instruments is the deficit of the resolvable overtones, since they attenuate faster than the fundamental frequency \cite{chaigne1991,chaigne1994}.   The current coupling scheme suffers from inferior energy transferring mechanism.  A physical limit, completely unaware to the musical population, exists to hinder the progress in the violin making.  A new acoustic mechanism to improve the couplings is much needed to transmit the energy efficiently.  

In fact, the problem lies in that there is considerable energy draining to the rim of the body through the tailpiece, the tailgut, and the endpin.  Here, we propose a functional acoustic insulation added between the strings and the tailpiece to reduce the energy loss to those accessories.  We fabricate the ball-shaped device with the diameter 12 millimeter and the weight less than 0.1 gram on the base of PU foam, called {\it String Bean\footnote{\begin{CJK*}{UTF8}{bkai} 琴弦四季豆 \end{CJK*}} }(SB), with the mechanical design for the functional purpose.  At the first glance, the spongy property of the PU foam might be thought to blur the sound quality.  On the contrary, our results demonstrate that the functionally-fabricated SB reduces the energy dissipation effectively in all frequency domain, resulting in enormous amplification of overtone intensity.  In this paper, we use a cello, as an example, to show that the proposed new acoustic mechanism promotes the violin family to a superior sounding instrument overall.  This physical method can be applied easily to the instruments of all sizes and ages, that could benefit not only professional musicians but also musical children.

\begin{figure}[!hbp]
\centering
\includegraphics[scale=0.12]{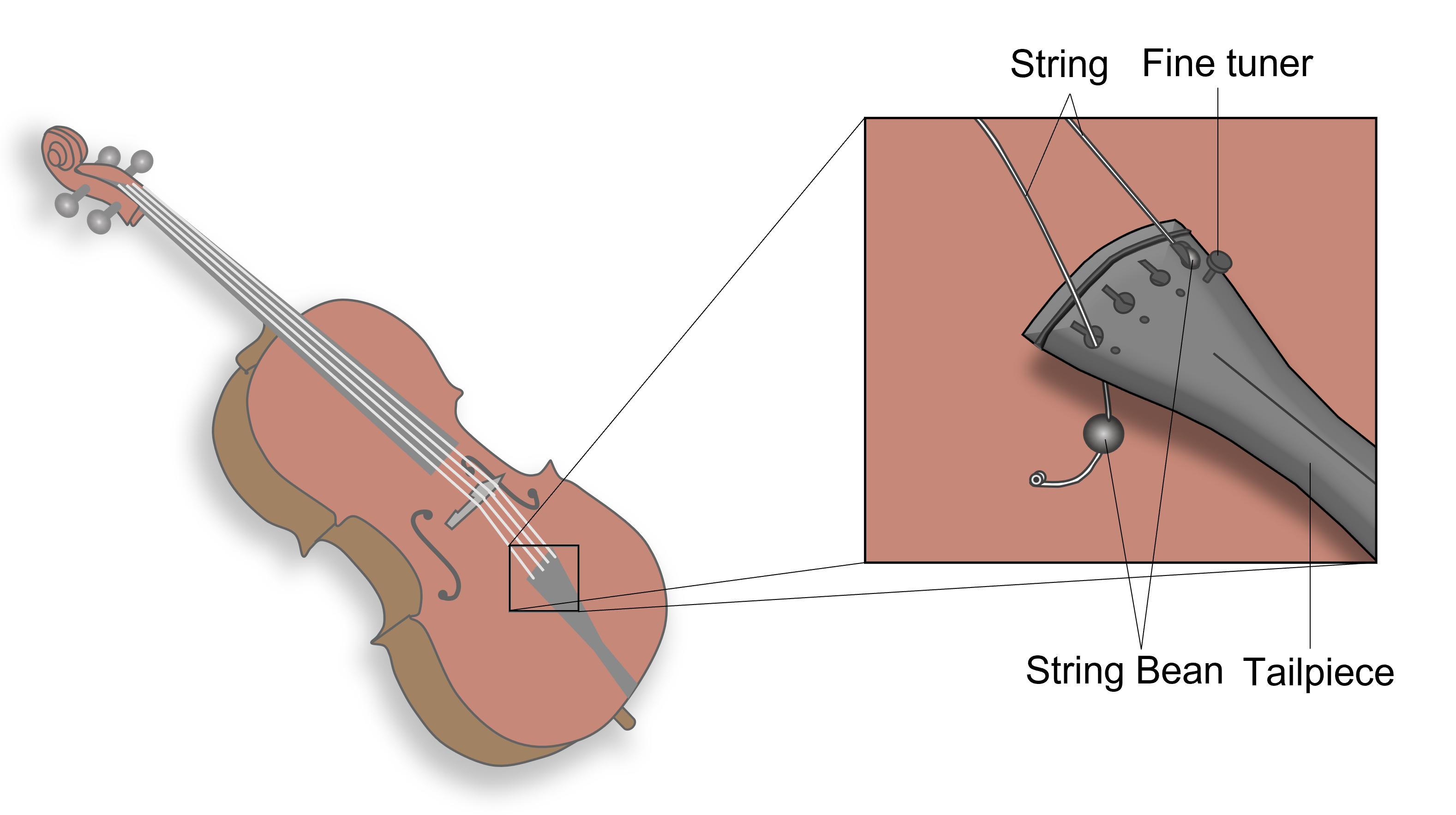}
\caption{{\bf The structural diagram of the SB installation in the violin family}.  The SB is sandwiched between the ball-end of the string and the accessories, so that the ball-end does not contact directly with the accessories.  In the inset, the structural diagram is illustrated that the SB is threaded in the string at the metal ball-end and mounted to the tailpiece in the normal way.  By the natural string tension, the SB is fixed in the fine tuner or under the tailpiece.  }
\label{fig1}
\end{figure}

\noindent
{\bf String bean installation and measurement}

The string threaded with SB at the metal ball-end is mounted in the normal way, fixing the SBs under the tailpiece or in the built-in fine tuner.   They are sandwiched between the string ball-end and the tailpiece/fine tuner naturally by the string tension as depicted in Fig. 1. The recording is conducted in a full anechoic chamber, including floor and ceiling.  The environmental noise is measured around 0.3 dB/Hz, corresponding to the background noise intensity 43.3 dB.  The plucking measurement is performed by a professional cello soloist (one of us, CCC) in the playing position on the 4 strings in the natural frequency $C_2$(65.4 Hz), $G_2$(98 Hz), $D_3$(146.8 Hz), $ A_3$(220 Hz) using the standard pizzicato technique.  10 samples of single pluck data are collected for each string without SB, and the same procedures are repeated after 4 SBs are installed in all the strings in the same cello.  A directional microphone is always kept 29 cm away from the bridge tip normal to the surface plate.  The data collection was taken in three hours successively.  The measurement of calibration reads $0.074\sin(2\pi f_0t)$  Pa = 85.7 dB SPL at $f_0=1000$ Hz, where SPL is the sound pressure level.  The high precision fast Fourier transformation is applied for the spectral analysis using the Python programming package {\it librosa} \cite{brian_mcfee_2022_6097378}.

\vspace{0.5cm}
\noindent
{\bf Signal analysis}

Fig. 2a shows the total intensity (dB SPL) of the single pluck data as the function of time (second) of the A string in the no SB (nSB) and the SB cases.  A fast ramp-up in the intensity is followed by an oscillatory exponential decay.  The oscillatory decay, instead of monotonic, indicates a coupled oscillation system \cite{halliday}.  The data of different strings exhibit similar behavior with different oscillatory patterns.  The signals consist of three components: (1) the \emph{desired} string tonal sound of the fundamental frequency and the overtones, (2) the persistent \emph{undesired} sound from non-resonant shaking of the accessories and the instrument body, and (3) the background noise.  The decay is followed by the floor-like hybrid of the undesired sound and the background noise.  

\begin{figure}[!hbp]
\center
\includegraphics[scale=0.38]{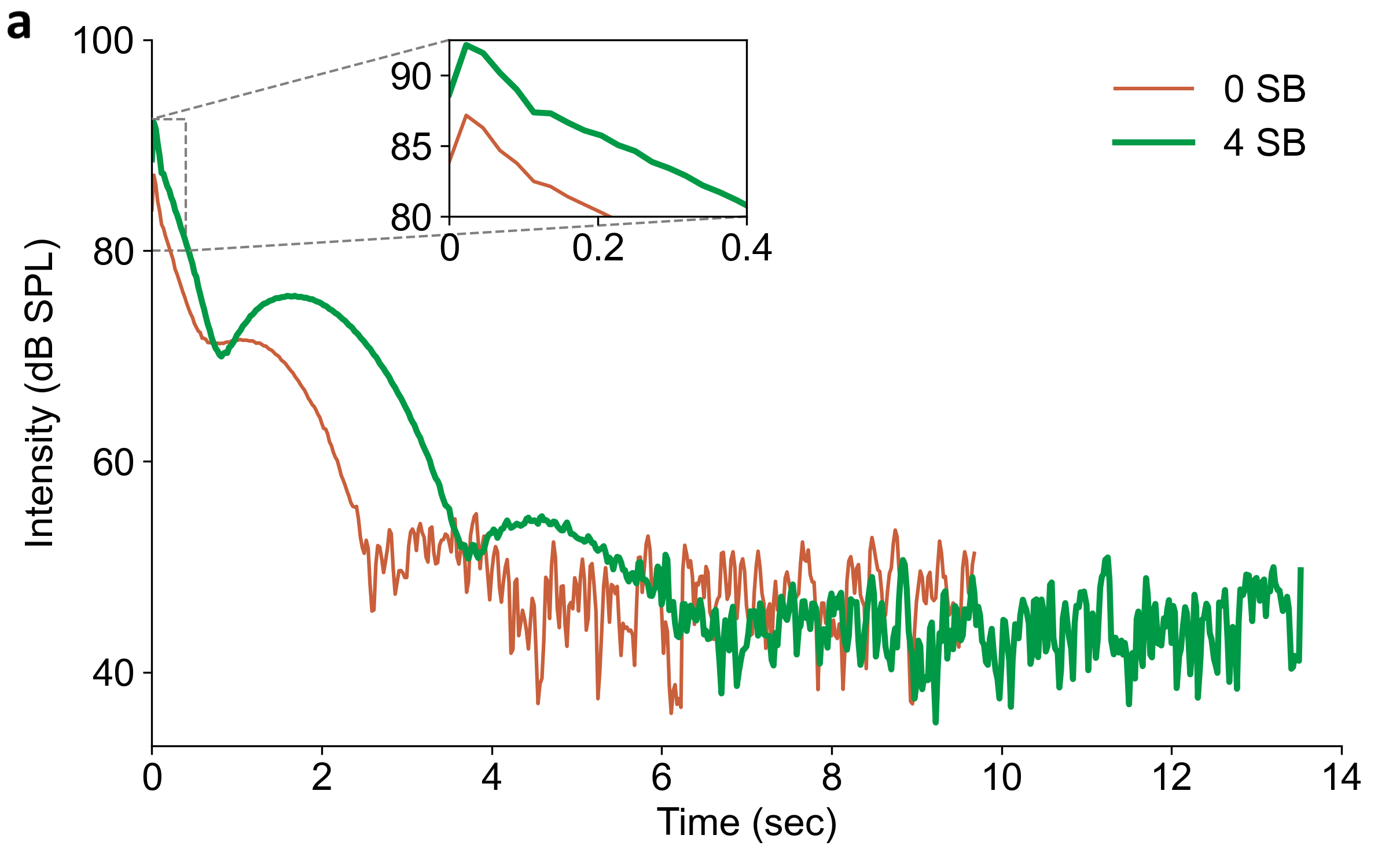}
\includegraphics[scale=0.38]{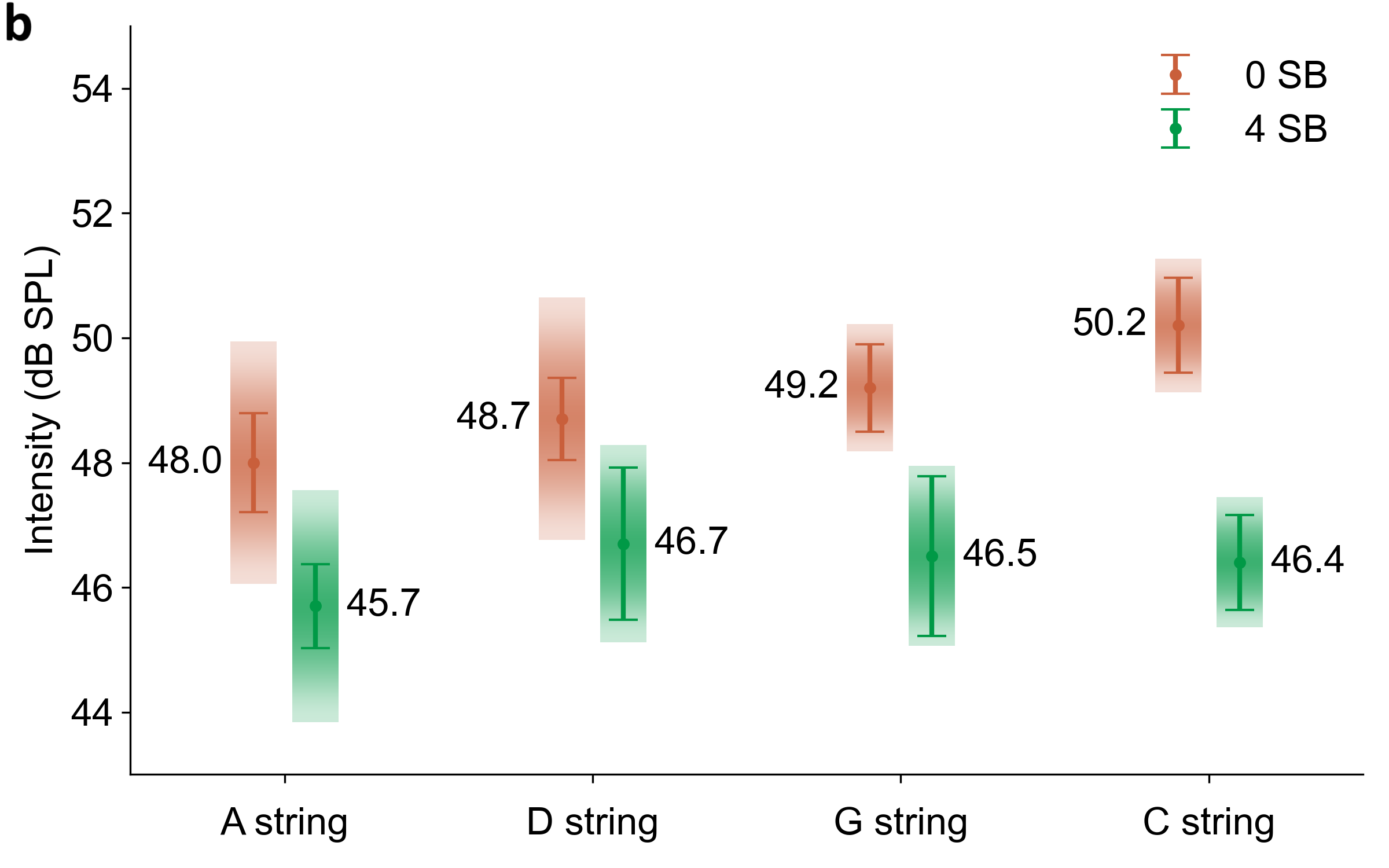}
\includegraphics[scale=0.38]{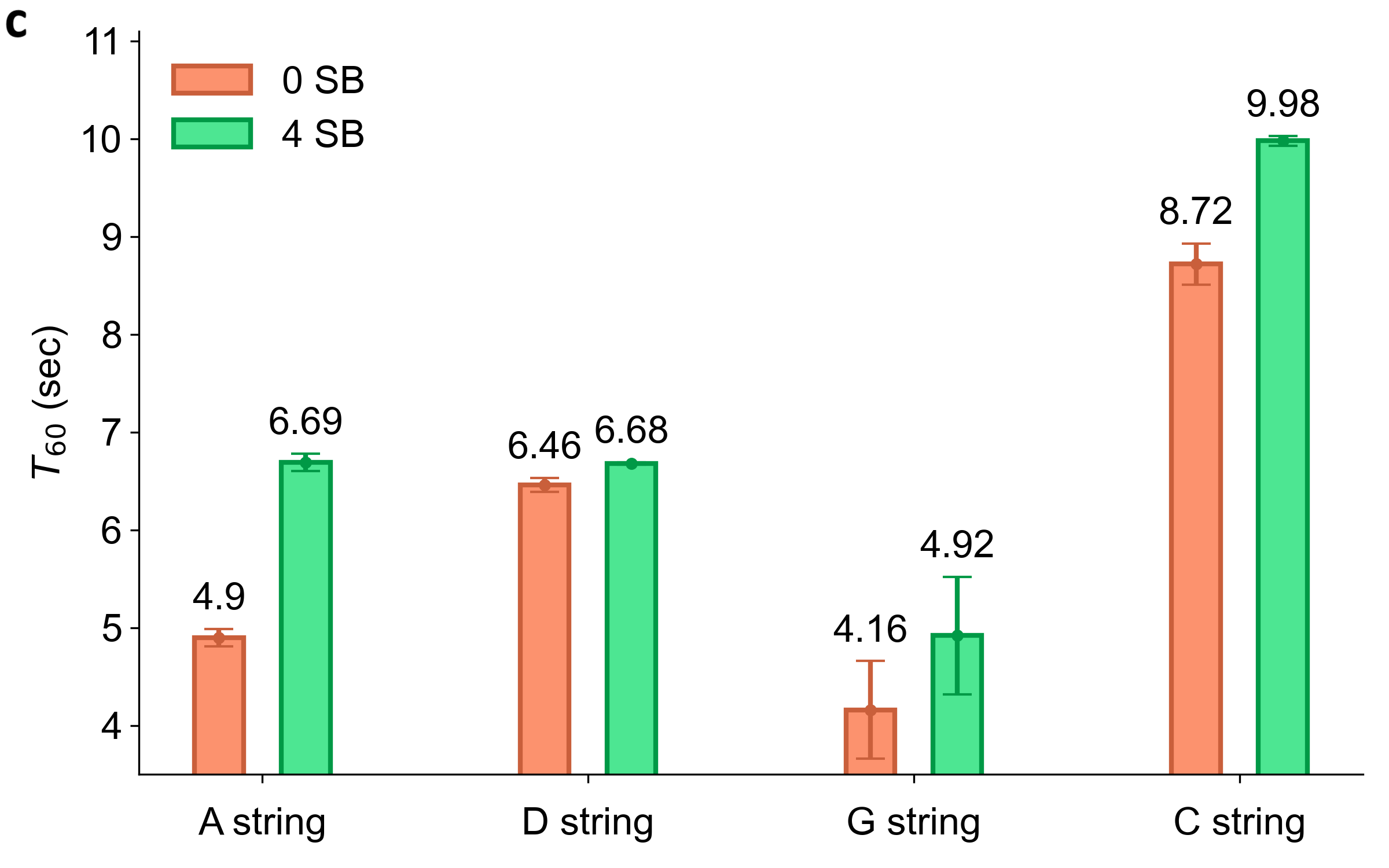}
\includegraphics[scale=0.38]{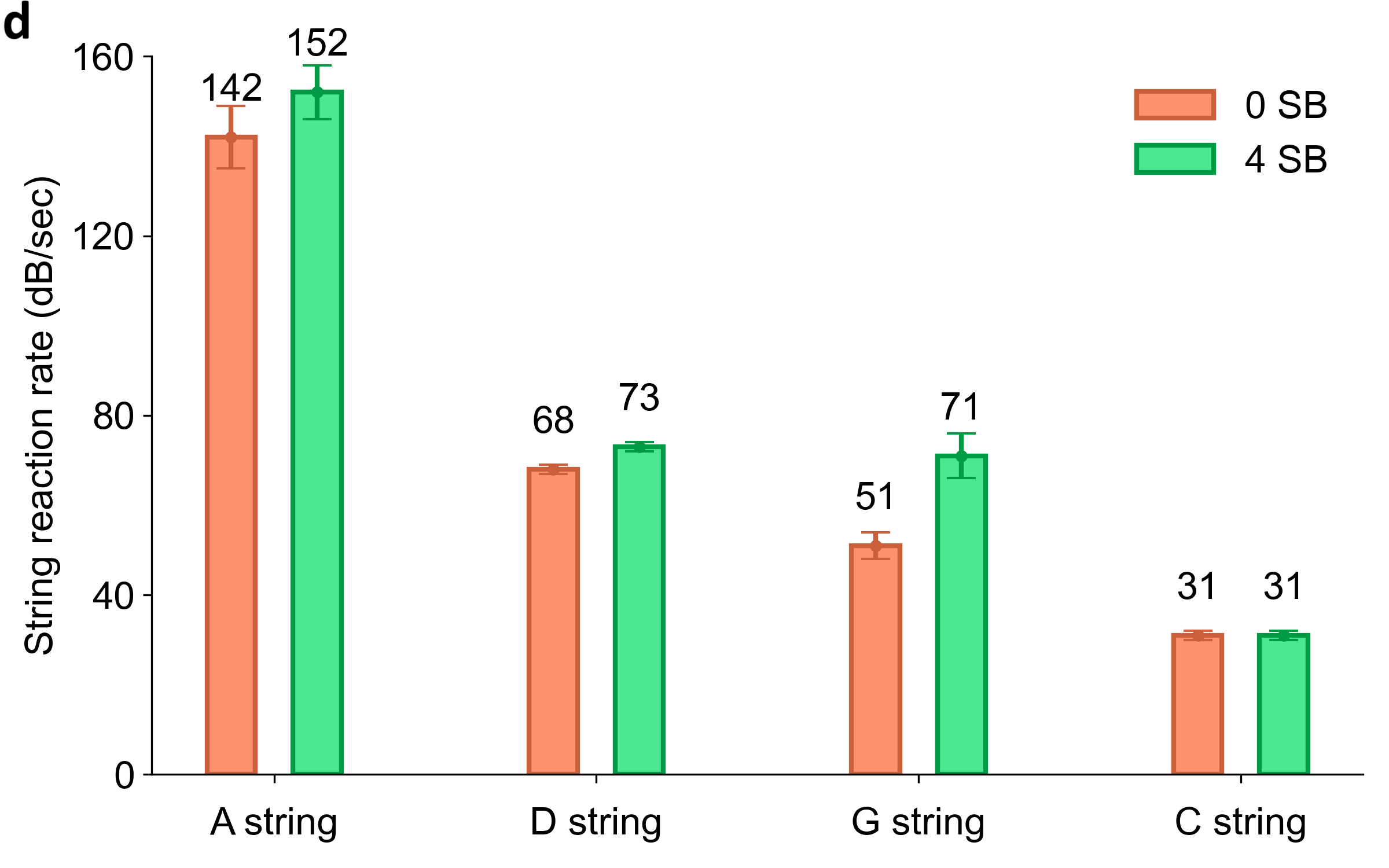}
\caption{{\bf Signal analysis of the single pluck on the cello}.  {\bf a}. The intensity of single pluck on the A string in the nSB (orange) and the SB (green) cases.  The inset shows the initial ramps up in the intensity associated with the string reaction rate (SSR).  The peak is followed by an oscillatory exponential decay and merges into the noise-like zone.  {\bf b}. The noise level and range in all strings.  The bar length indicates the noise range.  The standard deviation in the bar is obtained by averaging over the 10 samples of the noise level.  The noise level is lower in the SB cases for all strings.  {\bf c}. The $T_{60}$.  The violin family is a complicate coupled system, where $T_{60}$ is distinct in different strings.  {\bf d}. The string reaction rate (SSR) for all strings.  The SSR increases noticeably for the A, D, and G strings. }
\label{fig2}
\end{figure}

Several acoustic informations can be extracted in Fig. 2a.  First, the average intensity of the noise-like zone is computed to analyze the undesired sound.  Second, the reverberation time $T_{60}$, defined by the time interval of the 60 dB reduction from the peak, encodes the dissipation properties of the string signal.  Third, the slope of the initial ramp-up is associated with the responsiveness, namely the energy transferring rate from the strings to the plates.  Defining the slope (dB/second) as the string reaction rate (SSR), the larger SSR corresponds to the shorter reaction time of sounding.  Finally, the dynamic properties of the overtones are analyzed using the fast Fourier transformation.

In Fig. 2b, the average intensity in the noise-like zone is illustrated for all strings of the nSB and the SB cases over time interval $T=4$ seconds.  Different intervals of integration is adopted for different cases to secure meaningful noise-like information beyond the background noise. For example, on the $A$ string in the Fig. 2a, $T$ is taken from 5 to 9 seconds in the nSB case and from 8 to 12 seconds in the SB case.    The bars in the Fig. 2b represent the dB range of the noise-like zone.  All results include the standard deviation from the statistical averaging.  In the SB cases in Fig. 2b , the noise level decreases by 2.2 dB (A string), 2.4 dB (D), 2.7 dB (G), and 3.8 dB (C), and the level distribution becomes more uniform over the strings, as well.  As the background noise remains unaltered, the results suggest that the instrumental noise is highly suppressed.  The noise suppression manifests the dissipation reduction.  In other words, the instrumental dissipation intensity drops 61\% off (A string), 60\% off (D), 63\% off (G), and 73\% off (C) in the SB cases.  

The dissipation reduction implies that more mechanical energy is preserved and transmit to resonate the plates.  In this case, the vibration amplitudes of the plates should increase correspondently.  Averaging the peak intensity, we indeed observe the peak gain by 3.3$\pm 1.0$ dB (A), 4.7$\pm 0.6$ dB (D), 2.5$\pm 0.6$ dB (G), and 3.5$\pm 0.4$ dB (C) in the SB cases.  Considering that two (three) cellos playing simultaneously increase volume by 3.01 dB (4.77 dB) in comparison to a solo cello, the intensity gain in the SB cases is a significant difference.  The underlying mechanism of the profound effect will be elaborated in detail in the later section.  

In Fig. 2c, we compute the reverberation time $T_{60}$ by extrapolation.  First, we read out directly the time interval of the 60 deduction from the peak in the {\it natural logarithm scale} and obtain $T_{60log_{10}e}$, where $e$ is the Euler's number.  Then, $T_{60}$ is extrapolated by $T_{60log_{10}e}\ln 10$.   As shown in Fig. 2c, the $T_{60}$ increases by 1.79 seconds (A string), 0.22 seconds (D), 0.76 seconds (G), and 1.26 seconds (C) in the SB cases. In the full anechoic chamber, the reverberation time can represent the half-life time of instrument sounding.  The longer $T_{60}$ also supports the enhanced preservation of the mechanical energy.

Next, let us discuss the responsiveness.  In the Fig. 2d, the SSR is shown in positive correlation with the string tension in both the nSB and the SB cases.  It increases by 10 dB/second (A), 5 dB/second (D), 20 dB/second (G) in the SB cases.  Since SSR is inversely proportional to the reaction time, the results imply that the instrument becomes more responsive and is of better playability in the modified acoustic couplings.  Possibly due to the low tension, the SSR in the C string does not increases noticeably over the error bar.  A different mechanical specialization of SB might be needed for better responsiveness in C string.

In brief summary, the dissipation intensity acquires substantial suppression in the application of the SB.  The peak amplification, the longer reverberation time, and the shorter reaction time are the profound signatures for a sounding system of lesser dissipation.  Furthermore, the better intensity ratio of the higher string signal to the lower residual noise implies the enhancement of the sound projection.  To our best knowledge, it is the first time in the music history that the energy efficiency is seriously studied in manifestation of the significant improvement.

\vspace{0.5cm}
\noindent
{\bf Spectral analysis}
\begin{figure}[!hbp]
\center
\includegraphics[scale=0.38]{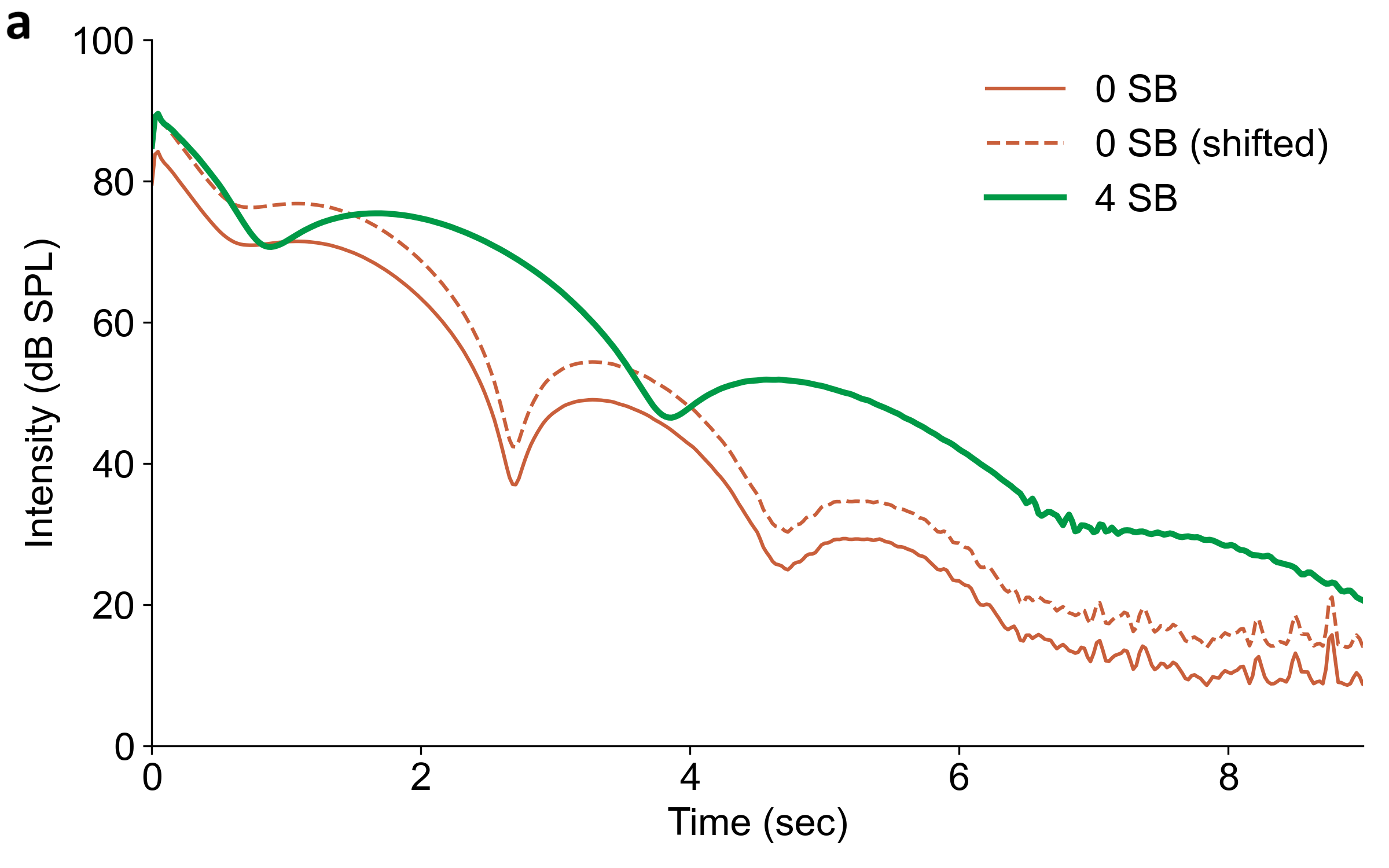}
\includegraphics[scale=0.38]{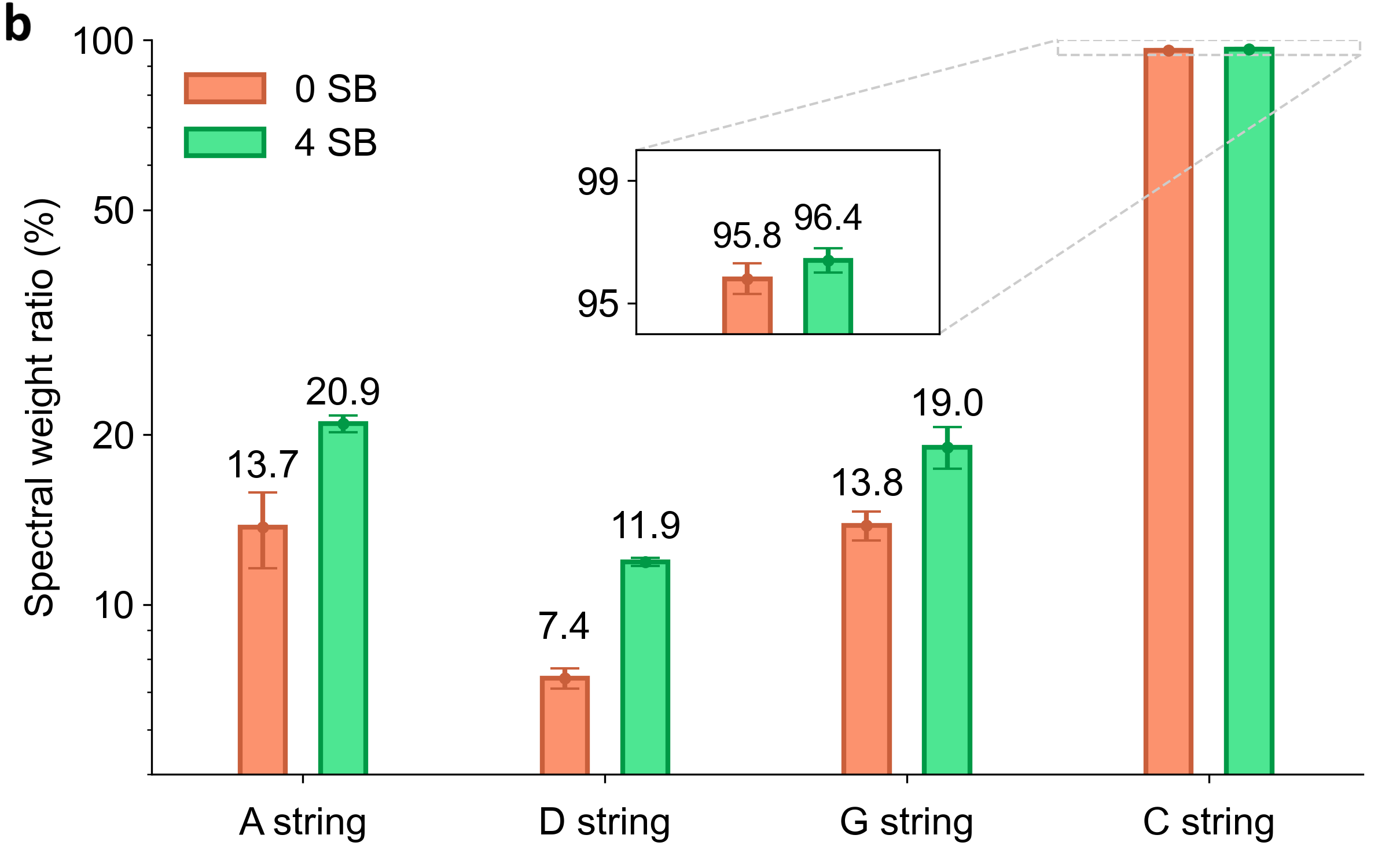}
\includegraphics[scale=0.34]{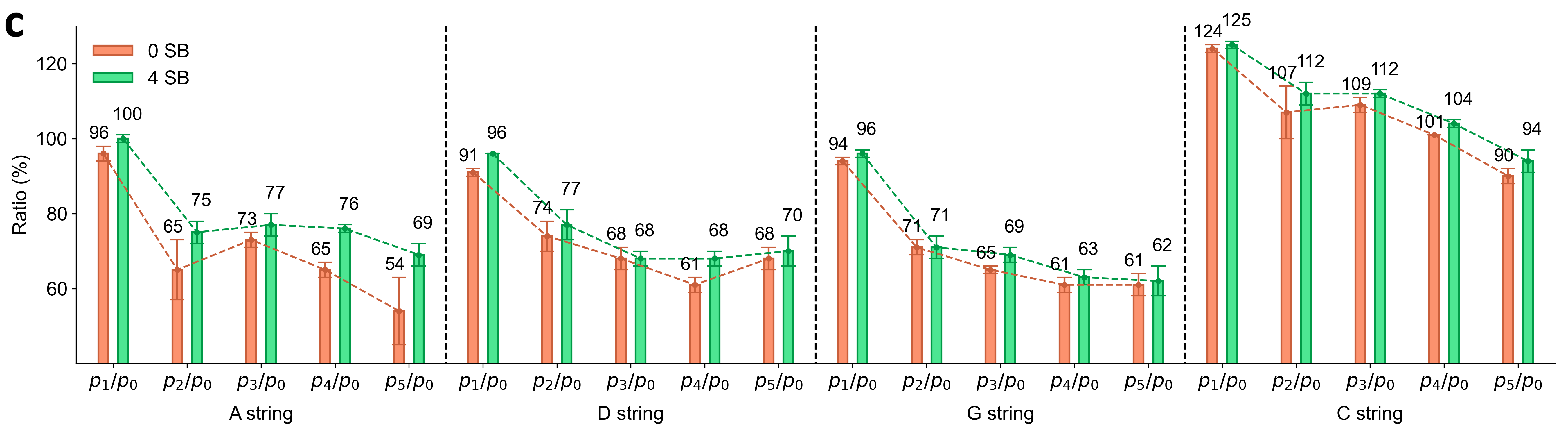}
\includegraphics[scale=0.34]{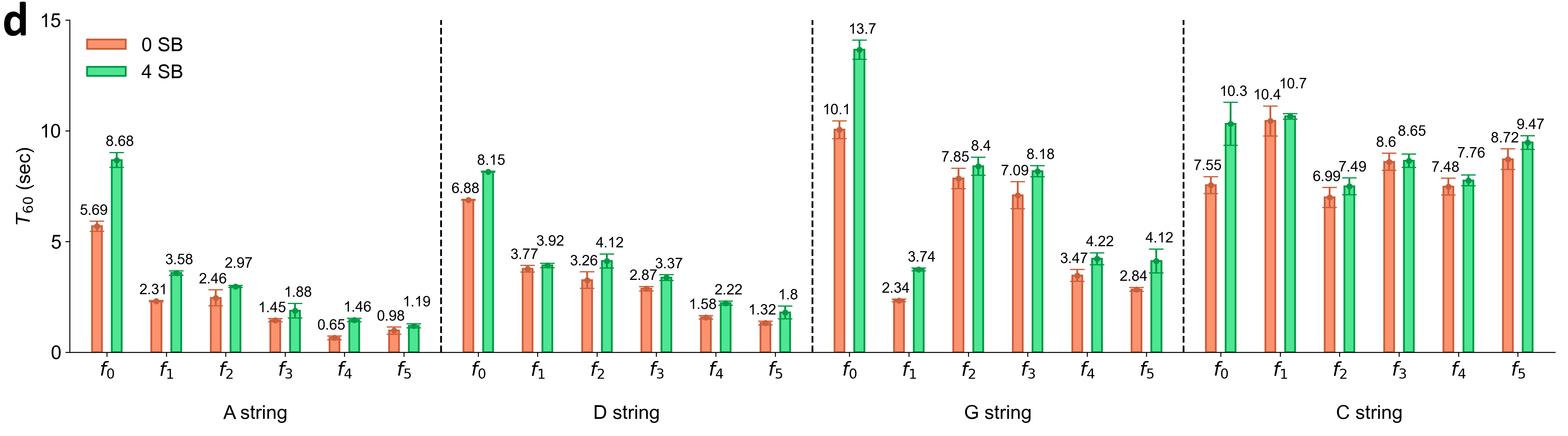}
\caption{{\bf Spectral analysis of the single pluck on the cello}. {\bf a}.  The intensity of the fundamental frequency $f_0$ of the A string in the nSB (orange) and SB (green) cases.  The dot line is the shifted data of the nSB case by a constant to match the peak in the SB case for the better visualization of the longer reverberation time in the SB case.  {\bf b}. The spectral weight of the overtones in the logarithmic scale.  {\bf c}. The ratio of $f_i/f_0$ measured at the peak of $f_0$ for $i=$1 to 5.  $p_i$ are the intensity of $f_i$ at the moment of $p_0$ the peak intensity of $f_0$.  The dot line indicates the spectral profile.  {\bf d}.  The reverberation time for all overtones in all strings.}
\label{fig3}
\end{figure}

In the following, we present the results of the spectral analysis using the high precision fast Fourier transformation.  A normal human ear can resolve up to 5 harmonics in a complex tone \cite{plomp1964, plomp1968, moore1995, moore2011}, although some study revealed the trained musicians can resolve more, up to 7 harmonics \cite{soderquist1970}.  Here, we show the results of the 6 harmonics, $f_0$ to $f_5$, where $f_0$ is the fundamental frequency and $f_i$, $i=$ 1 to 5, are the overtones.  In A string, $f_0$ is $A_3$ (220 Hz), $f_1$ is $A_4$ (440 Hz), $f_2$ is $E_5$ (660 Hz), and so on.  The dynamics of the harmonics, for example, in Fig. 3a shows the $f_0$ of the A string in the nSB and the SB cases extracted from Fig. 2a.  In the spectral analysis, we are interested in (1) the $T_{60}$ of the harmonics, (2) the relative intensity of the overtone to $f_0$, and (3) the spectral weight $\mathcal{W}_5$ defined by
\begin{eqnarray}
\mathcal{W}_5=\frac{\sum_{i=1}^5 \mathcal{E}_i}{\mathcal{E}_{total}}, \label{spw}
\end{eqnarray}
where $\mathcal{E}_i$ is the $i^{th}$ overtone energy and $\mathcal{E}_{total}$ is the total energy of the signal.  The spectral weight decodes the overtone components in the total signal.

Fig. 3b shows the results of $\mathcal{W}_5$ as 13.7\% (A), 7.4\% (D), 13.8\% (G), and 95.8\% (C) in the nSB cases and 20.9\% (A), 11.9\% (D), 19.0\% (G), and 96.4\% (C) in the SB cases.  All strings exhibit substantial increase in the overtone components.  In particular, the spectral weight in the D string of the SB case boosts to 161\% of the one in the nSB cases.  The 0.6\% increment of $\mathcal{W}_5$ in the C string is highly salient, since the ratio $\mathcal{E}_0/\mathcal{E}_{total}$ is only 0.8\% in the nSB case, where $\mathcal{E}_0$ is the $f_0$ energy in the C string.  In connection to perceivable hearing, we compute the intensity ratios of the overtone to the $f_0$. Since all harmonics are dynamical, we compute $p_i/p_0$, where $p_i$ are the intensity of the $i^{th}$ harmonics at the moment of the $f_0$ peak. In the Fig. 3c, most of the $p_i/p_0$ for $i=1$ to $5$ exhibit substantial boost, in particular 15\% more in the 5$^{th}$ overtone of the A string and 4.3\% in average of all.

Finally, the $T_{60}$ of the overtones is extracted by fitting the linear slope numerically within 5\% the asymptotic standard error.  Fig. 3d shows the $T_{60}$ of all 6 harmonics from $f_0$ to $f_5$ in the nSB and the SB cases.  All six harmonics demonstrate slower decay in the SB cases.  Overall, the $T_{60}$ increases by 29\% in average.  In summary, among the three components in the sound signal mentioned above, the new acoustic couplings effectively enhance the \emph{desired} string harmonic sound and suppress the persistent \emph{undesired} sound from non-resonant shaking of the accessories.  Those results indicate the enhancement of the harmonic resolvability in manifestation of the timbre complexity. 

\vspace{0.5cm}
\noindent
{\bf The physical modeling}

The underlying mechanism of sounding in the violin family can be easily understood as the following.  The violin family consists of two oscillation systems, the strings and the wooden plates, coupled by the accessories.  The strings are the first oscillation system that receives the energy from the player.  After the reaction time, the energy transmits thoroughly to resonate the wooden plates.  In the ideal case, the wooden plates should acquire all the mechanical energy from the strings.  However, in the real situation, the energy is partly absorbed by the accessories and other non-resonant parts of the body.  They are shaken along with the strings and the plates.  Their non-resonant vibrations also make sound.  Unlike the resonant sound from the plates, those undesired sounds does not propagate far in distance.  The string harmonic signals propagates distantly in space, since they are in resonance of the large-scaled collective motion of plate vibrations.

\begin{figure}[!hbp]
\center
\includegraphics[scale=0.5]{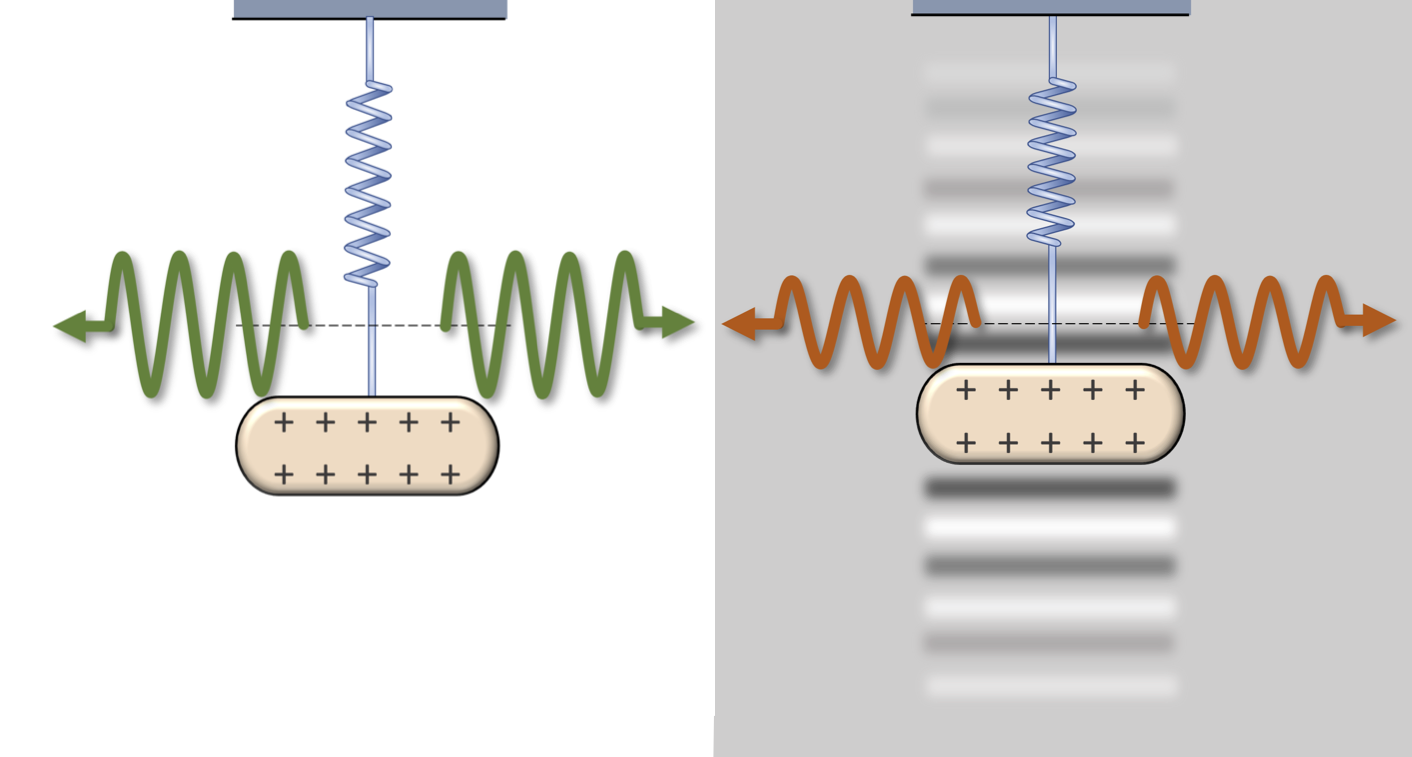}
\caption{{\bf A simple physical model explaining the sounding behavior of the violin family.}  The charged spring system radiates the electromagnetic wave as it oscillates.  In the ideal vacuum situation (left panel), all the mechanical energy transfers to the electromagnetic energy (green wavy line).  In the viscous media (right panel), the mechanical energy dissipates to generate the attenuating wave, and the amplitude of the oscillation shrinks quicker, resulting in lower power and shorter half-life time of the EM radiation decay (orange wavy line).}
\label{fig4}
\end{figure}

The results of our analysis can be easily explained by a simple physical model of the charged spring system as depicted in Fig. 4 \cite{halliday}.  As the spring with the charged massive object oscillates, electromagnetic radiation (EM) is generated because of the charge acceleration.  The radiation power averaged in a cycle can be computed approximately as \cite{larmor1897}
\begin{eqnarray}
\mathcal{P}_{\small \tt EM}=\alpha_{\small \tt EM} x_c^2 \label{larmor},
\end{eqnarray}
where $\alpha_{\tt EM}$ is the frequency-dependent coefficient and $x_c$ is the amplitude of the spring in the cycle.  The sound emission in the violin family follows the same relation of the plate vibration amplitudes with different coefficients.  The exponential decay of the electromagnetic radiation can be derived from the Eq. 2.  In the ideal case of vacuum, all the mechanical energy in the spring eventually transfers to the electromagnetic energy.  However, in the presence of the viscous media, the attenuating wave is also generated as shown in the Fig. 4.  The sounding behavior of the ordinary violin family is in analogy to the charged spring system in the viscous media, where the attenuating wave represents the undesired sounds from the non-resonant parts of the body and the accessories, and the electromagnetic radiation represents the sound of the harmonics.   The results in the Fig. 2b indicates that the new acoustic couplings with the SB suppress the non-resonant shaking.   In other words, the cello with the SB becomes less dissipative similar to the spring system in vacuum with a longer half-life time.  Concomitantly, the longer reverberation time $T_{60}$ in the SB cases in the Fig. 2c and Fig. 3d is in intimate relation to the dissipation reduction in the Fig. 2b.  

In this model, it is easy to understand the intensity amplification observed in the Fig. 2a and the Fig. 3a.  Transferring the mechanical energy from the strings to the wooden plates is equivalent to triggering the oscillation in the charged spring system.  Due to the suppression of the non-resonant shaking in the SB cases, more energy is transferred to the plates, corresponding to a larger initial amplitude $x_c$ of the oscillation in the charged spring system.  According to Eq. 2, the larger initial $x_c$ implies higher peak intensity.   In other words, the amelioration of the acoustic couplings enhances the intensity of the harmonic signal, achieving higher efficiency in energy conversion process from the biological to the sound. 

\vspace{0.5cm}
\noindent
{\bf Conclusion}

In the traditional knowledge of the violin making, the SB might be regarded as a wrong thing put in a wrong place.  On the contrary, our approach systematically improves the efficiency in the energy transmission, that completely changes the understanding in acoustics of the musical instruments.  The violin family becomes a better sounding instrument extensively in the ameliorated acoustic couplings, exhibiting the substantial enhancement of the sound projection and the overtone components.  In addition, its ultralight feature manifests itself to work silently to release the sounding potential unknown to people for centuries.   Shedding light on the advance of human's aspiration for better sound quality, the application of SB in string-based instruments opens up the new research direction as well as the new phase of the sound quality in the music history.


\section*{Acknowledgement}
CHC deeply appreciates the stimulating discussions with the violinists Ray Chen, Chinn-Horng Chen, Chien-Tai Hsu, I-Ching Li, Shien-Ta Su, Yu-Chien (Benny) Tseng, Ting-Yuh Wu, and the luthiers David Lien, Ting-Ting Wang.  LS is supported in part of the Thematic Research Program AS-TP-111-M02 by Academia Sinica and in the 109-2221-E-001-018 -MY3 and the 110-2221-E-001-010-MY3 by Ministry of Science and Technology in Taiwan.

\section*{Author contributions}
LS performed the calculation, including the high precision fast Fourier transformation in Python, analyzed the data, and prepared all the figures in the manuscript.  CCC plucked the cello.  CHC contributed to the fabrication and the mechanical design of the String Bean, the physical model building, the data interpretation, and the manuscript writing and editing.  YML contributed to the fabrication of the String Bean, discussion on the results, and the manuscript proofreading.  LCL collected and sorted the data in the recording.  YWL coordinated the recording and performed some early test analysis in Matlab.  MSB offers the full anechoic chamber under the maintenance of his group. 

\section*{Competing interests}
CHC currently holds the patent in United States of America (No. 11205406) and in South Korea (10-2020-0039584) associated with the functional locations where the String Bean is installed and declares the patent pending in European Union (EP3882904A2) and China (CN113554995A).  Other authors declare no competing interests



\end{document}